\documentclass[12pt]{article}
\pdfoutput=1
\usepackage{graphicx}
\usepackage{epsfig}
\usepackage{amsmath,amssymb,hyperref,enumerate}
\usepackage{pstricks}
\usepackage{slashed}
\usepackage{amsfonts}
\usepackage{url}
\usepackage{color}
\usepackage{appendix}
\usepackage{bm}
\usepackage{subfig}
\usepackage{cite}
\hypersetup{colorlinks,citecolor= black,linkcolor= black}
\definecolor{nicered}{rgb}{0.7,0.1,0.1}
\definecolor{nicegreen}{rgb}{0.1,0.5,0.1}

\usepackage{setspace}

\allowdisplaybreaks

\def\({\left(}
\def\){\right)}
\def\[{\left[}
\def\]{\right]}

\usepackage[margin=1.in]{geometry}

\begin{document}


\onehalfspacing

\begin{flushright}
OSU-HEP-14-09\\
UMD-PP-014-012
\end{flushright}

\renewcommand{\thefootnote}{\fnsymbol{footnote}}

\begin{center}
\LARGE {\bf \Large Determining Majorana Nature of Neutrino from\\[-0.1in]
Nucleon Decays and \boldmath${n-\bar{n}}$ oscillations}
\end{center}

\vspace{0.5cm}

\begin{center}
{\large  K.S. Babu$^a$\footnote{Email:
babu@okstate.edu} and Rabindra N. Mohapatra$^b$\footnote{Email: rmohapat@physics.umd.edu}}

\end{center}

\begin{center}
\it $^a$Department of Physics, Oklahoma State University,
Stillwater, Oklahoma 74078, USA

\bigskip
\it $^b$Maryland Center for Fundamental Physics and Department of Physics\\
 University of Maryland, College Park, MD 20742, USA
\end{center}

\renewcommand{\thefootnote}{\arabic{footnote}}
\setcounter{footnote}{0}

\bigskip

\begin{abstract}

We show that discovery of baryon number violation in two processes with at least one obeying the selection rule $\Delta (B-L) = \pm 2$ can determine
the Majorana character of neutrinos.  Thus observing  $p \rightarrow e^+ \pi^0$ and $n \rightarrow e^- \pi^0$ decays, or
$p \rightarrow e^+ \pi^0$  and $n-\bar{n}$ oscillations, or $n \rightarrow e^- \pi^+$  and $n-\bar{n}$ oscillations
would establish that neutrinos are Majorana particles.  We discuss this in a model-independent effective operator approach.

\end{abstract}

An outstanding question in particle physics today is whether neutrinos are their own antiparticles (Majorana fermions) or not.
This  can be settled most directly by the detection of a nonzero signal in searches for neutrinoless double beta decay \cite{review},
which would establish the Majorana character of the electron type neutrino. If the neutrino masses exhibit a normal and non-degenerate
mass hierarchy, however, the prospects are bleak for observing this decay in the near future as it would require experiments with much higher sensitivity than currently available. It may therefore be of interest to search for alternative ways to probe the Majorana character of neutrinos, which plays a pivotal role in the understanding of the physics behind neutrino masses.

In this Letter we suggest that discovery of baryon number violation in two processes, with at least one obeying the selection rule
$\Delta (B-L) = \pm 2$ for the change in the number of baryons minus leptons, can provide such an alternative.  SuperKamiokane and
other deep underground experiments on the horizon are sensitive to nucleon decays of the type $p \rightarrow e^+ \pi^0$ which obeys
the selection rule $\Delta(B-L) =0$, as well as $n \rightarrow e^- \pi^+$, obeying $\Delta(B-L) = -2$ selection rule \cite{baryon}. These experiments
are also sensitive to matter disintegration caused by neutron--antineutron $(n-\bar{n})$ transition in nuclear matter, a
$\Delta(B-L) = -2$ process.  As we shall see, discovery of any two of these (or certain related) processes would establish the Majorana
character of the neutrino, which is intrinsically a $\Delta(B-L) = \pm 2$ effect.  To arrive at this conclusion we rely on a model-independent
effective operator analysis.  This connection between baryon number violation and Majorana character of the neutrino should
provide added impetus to higher sensitivity nucleon decay searches at the proposed LBNF facility \cite{LBNF} and the HyperKamiokande
experiment \cite{HyperK}. This should also provide added motivation for a high sensitivity search for $n-\bar{n}$ oscillation with free neutrons which is under consideration at the European Spallation Source (ESS) currently ~\cite{ess}.

We focus on the three baryon number violating processes listed below:
\begin{equation}
(i)~~ p \rightarrow e^+ \pi^0;~~~~~(ii)~~ n \rightarrow e^- \pi^+;~~~~~(iii)~~ n-\bar{n}~{\rm oscillations}~.
\label{decay}
\end{equation}
Process $(i)$ obeys $\Delta(B-L) =0$, while processes $(ii)$ and $(iii)$ obey $\Delta(B-L) = -2$. To establish the Majorana
nature of the neutrino from here, at least two of these  processes will have to be observed.  While discovery of nucleon decay
in itself will be profound, once a certain decay mode of the nucleon is seen, it is perhaps not unreasonable to expect additional
decay modes as well. This is what we assume in getting to our proof. But before proceeding, we note that the electron and the positron
in the final states of Eq. (\ref{decay}) can be replaced by  $\mu^\pm$.  The discovery of $p \rightarrow \mu^+ \pi^0$ and
$ n \rightarrow \mu^- \pi^+$ will establish the Majorana character of the muon type neutrino, something not possible in neutrinoless
double beta decay owing to kinematic limitations.  Furthermore, the pions Eq. (\ref{decay}) may be replaced by other mesons, including kaon, $\rho$,
$\omega$ etc, in which case our proof for the Majorana nature of neutrinos will still go through, aided by the dressing of diagrams
by the weak $W^\pm$ gauge boson which can change the strangeness quantum number.  Thus the more general baryon number
violating processes relevant for establishing Majorana nature of neutrinos
are $p \rightarrow \ell^+ M^0$; $n \rightarrow \ell^- M^+$; and $n-\bar{n}$ oscillations, with $\ell$ being $e$ or $\mu$ and $M$ being
any light meson.

\begin{figure}
\begin{center}
\includegraphics[width=7.5cm]{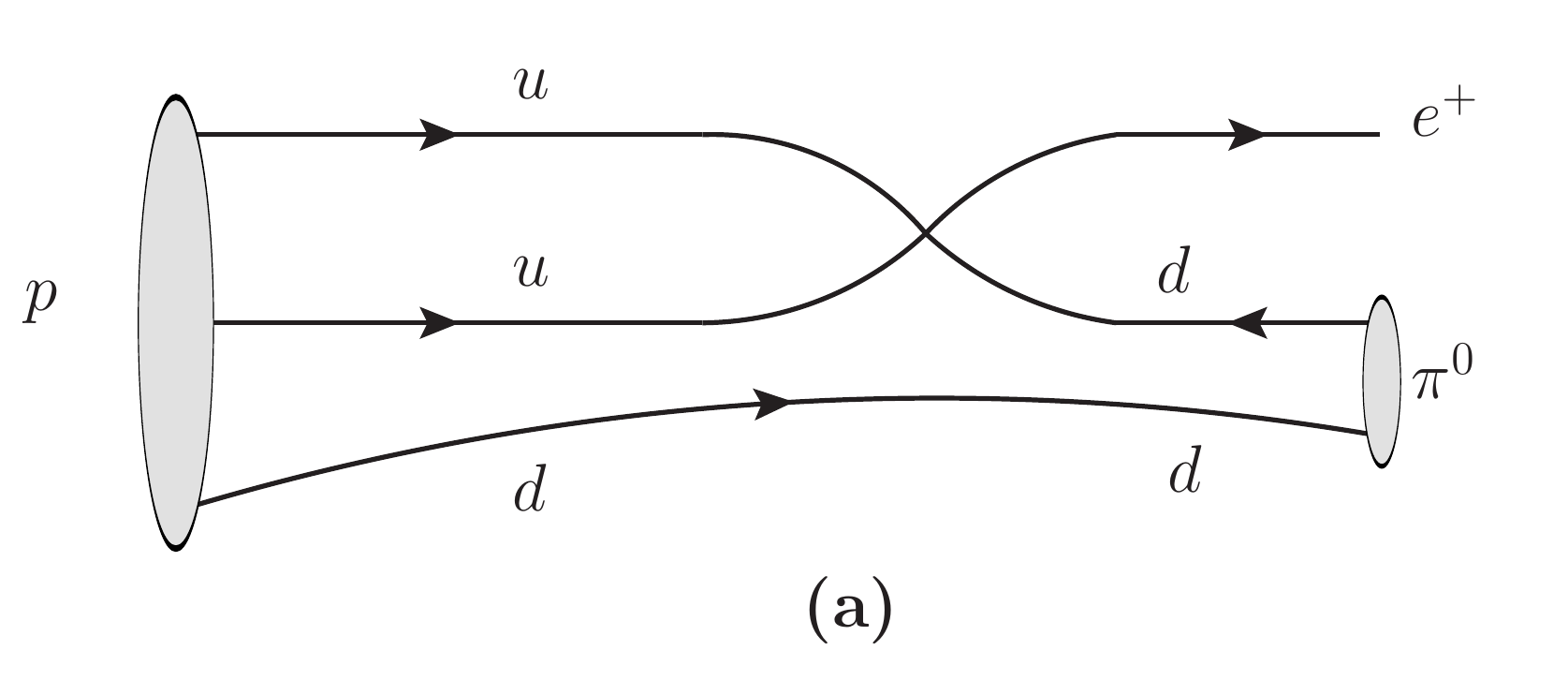}
\includegraphics[width=7.5cm]{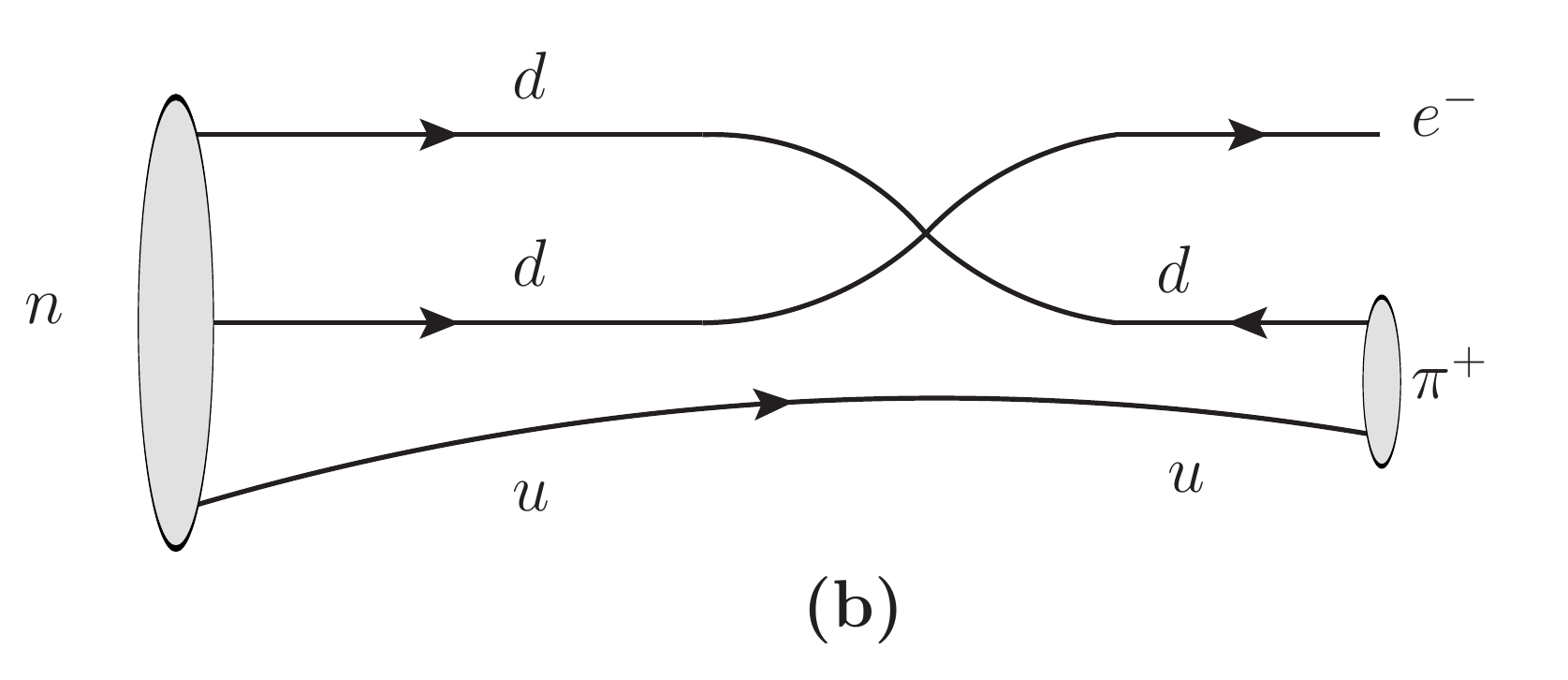}
\vspace*{1.0cm}
\includegraphics[width=7.5cm]{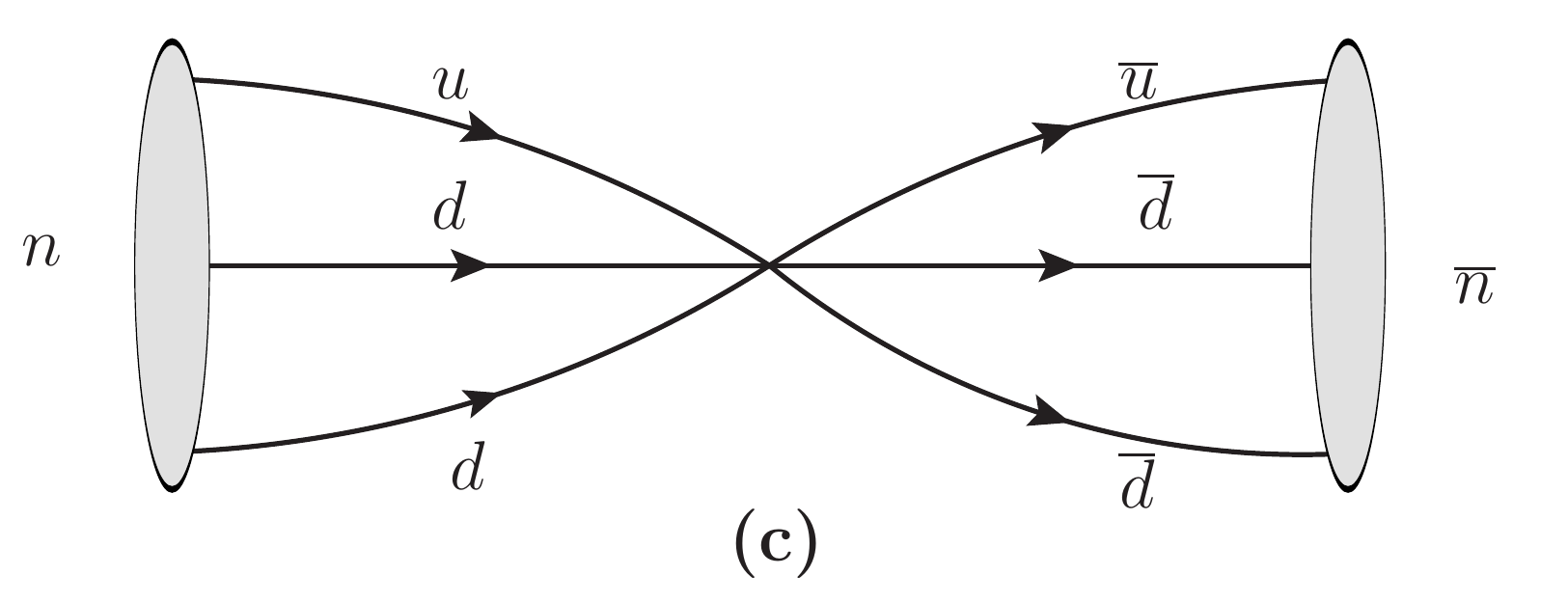}
\caption{Diagrams responsible for (a) $p \rightarrow e^+ \pi^0$ decay; (b) $n \rightarrow e^- \pi^+$ decay; and (c) $n-\bar{n}$ oscillations. }
\label{Fig1}
\end{center}
\end{figure}
In Fig. \ref{Fig1}  we depict the Feynman diagrams responsible for the $B$-violating processes listed in Eq. (\ref{decay}).  In a model-independent approach, we can characterize the $p \rightarrow e^+ \pi^0$  decay by the dimension six effective Lagrangian \cite{weinberg1}
\begin{eqnarray}
{\cal L}^{\rm eff}( p\rightarrow e^+ \pi^0) &=& \frac{1}{\Lambda^2_p} \left[c_1 \, (u_R d_R)(u_L e_L)
 + c_2 \,(u_L d_L)(u_R e_R) \right. \nonumber \\
&\,& \left. + c_{\{3,4\}}\, (u_L d_L)((u_L e_L)  + \,c_5 \, (u_Rd_R)(u_Re_R) \right]
\label{pdecay}
\end{eqnarray}
suppressed by two powers of the cutoff $\Lambda_p$.  Here color contraction, which is unique, should be understood.
Similarly the decay $n \rightarrow e^- \pi^+$ arises from the dimension seven
effective Lagrangian given by \cite{weinberg2,bm1}
\begin{eqnarray}
&\,& \hspace*{-0.5cm} {\cal L}^{\rm eff}( n\rightarrow e^- \pi^+)  = \frac{\left\langle H^0 \right \rangle}{\Lambda^3_n} \left[ \hat{c}_5 \, (d_R d_R)(d_L e^c_L) + \hat{c}_6 \,(d_R d_R)(d^c_L e_L)^*  \right] \nonumber \\
&+& \frac{1}{\Lambda_n^3} \left[ \hat{c}_7\, (\bar{e}_L \gamma_\mu d_L)(d^c_L \partial^\mu d^c_L)^*
+\hat{c}_8 \, (d^c_L \partial_\mu e_L)^*(\bar{d^c_L} \gamma^\mu d_L)
+ \hat{c}_9 \, (d^c_L \partial_\mu d^c_L)^*(\bar{d^c_L} \gamma^\mu e^c_L) \right].
\label{ndecay}
\end{eqnarray}
Note that these Lagrangian terms involve a vacuum expectation value of the Standard Model Higgs field, $\left\langle H^0 \right
\rangle \simeq 174$ GeV, or a derivative which would yield a light fermion mass in the decay amplitude, and thus are suppressed by
three powers of a cutoff scale $\Lambda_n$. (For applications of such $d=7$ terms to nucleon decay, baryogenesis and collider signals 
see Ref. \cite{bm1,barr}.)  Neutron-antineutron oscillations
arise from the dimension nine effective Lagrangian  suppressed by five powers of a cutoff scale $\Lambda_{n\bar{n}}$ and is
given by
\begin{equation}
{\cal L}^{\rm eff}(n-\bar{n}) = \frac{1}{\Lambda_{n\bar{n}}^5}\left[ c_1'\, (u_Rd_R)(u_R d_R)(d_R d_R) + ...\right].
\label{nnbar}
\end{equation}
Here there are a total of eighteen terms \cite{rao} which obey four constraint equations \cite{goran}.  For brevity we have
not displayed them all.  These terms are all similar to the term shown in Eq. (\ref{nnbar}), but differ in their
chiral structure, Lorentz contraction as well as color contraction. For our
purpose an illustrative term is sufficient.

The SuperKamiokande experiment has set a limit on the inverse decay rate for $p \rightarrow e^+ \pi^0$, $\tau(p \rightarrow e^+ \pi^0) > 1.4 \times 10^{34}$ yrs \cite{SK1}. This leads to a lower limit of
$\Lambda_p > 4.9 \times 10^{15}$ GeV, where we used the most recent lattice evaluation of the nucleon matrix element, $|\alpha_H| = 0.012$ GeV$^3$.
The current best limit on $n \rightarrow e^- \pi^0$ lifetime is $\tau(n \rightarrow e^- \pi^0)> 6.5 \times 10^{31}$ yrs from
IMB experiment \cite{IMB}, which results in the limit $\Lambda_n > 6.6 \times 10^{10}$ GeV.  
It would be interesting to see if SuperK can improve this limit with data already collected.
SuperKamiokande has also searched for
matter disintegration caused by $n-\bar{n}$ transition inside nuclear matter,
and has set a limit of $\tau > 1.89 \times 10^{32}$ yrs \cite{SK2},
which would correspond to $\Lambda_{n\bar{n}} > 10^5$ GeV, with a larger uncertainty resulting from the relevant
nuclear matrix element. An experiment carried out at ILL with free neutrons has set a limit on the
$n-\bar{n}$ oscillation time of $\tau(n-\bar{n}) > 0.86 \times 10^8$ sec \cite{ILL}, which would also yield comparable limit, $\Lambda_{n\bar{n}} > 10^5$ GeV.

\begin{figure}
\begin{center}
\includegraphics[width=4.5cm]{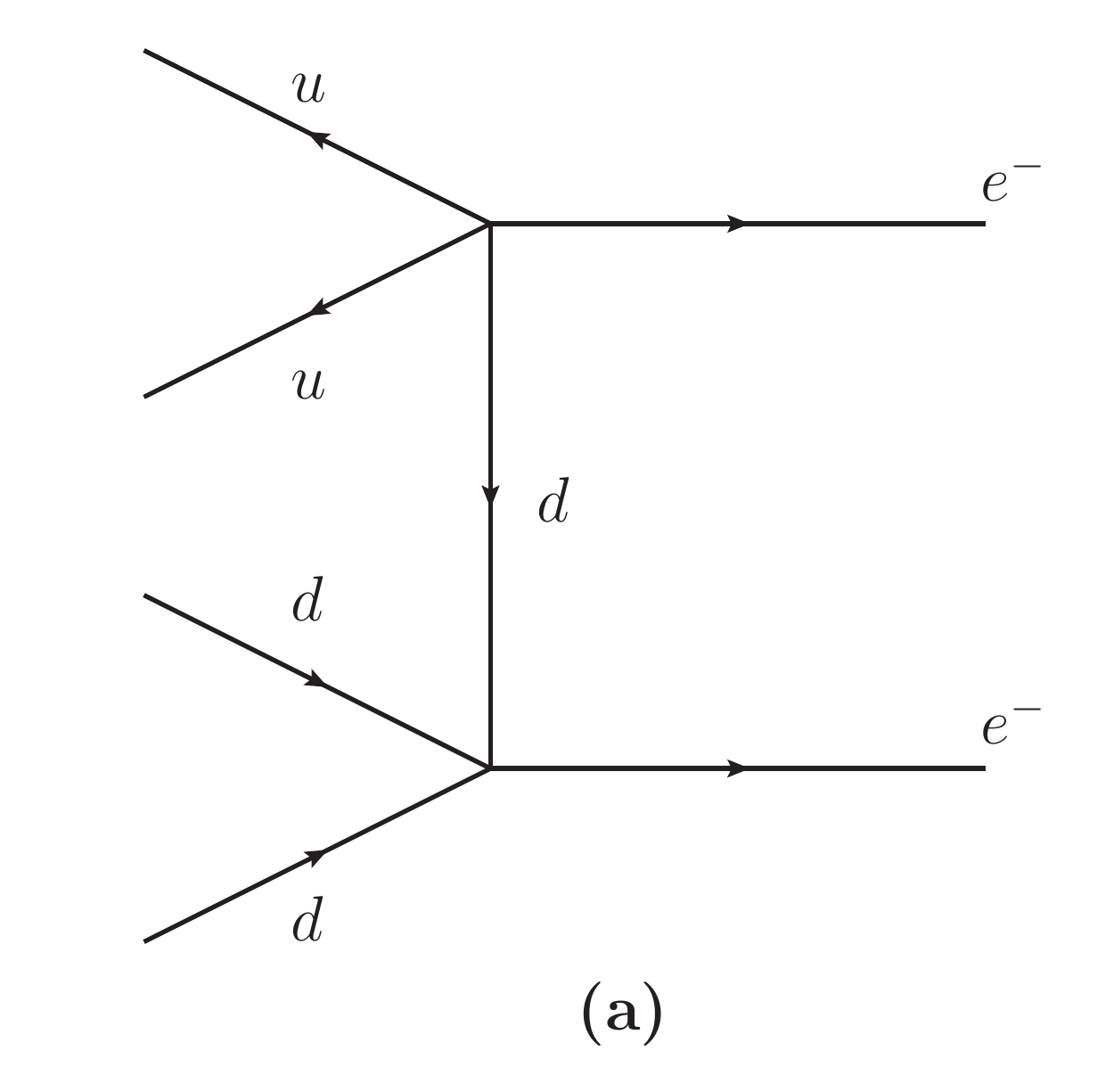}
\hspace*{0.3cm}
\includegraphics[width=4.5cm]{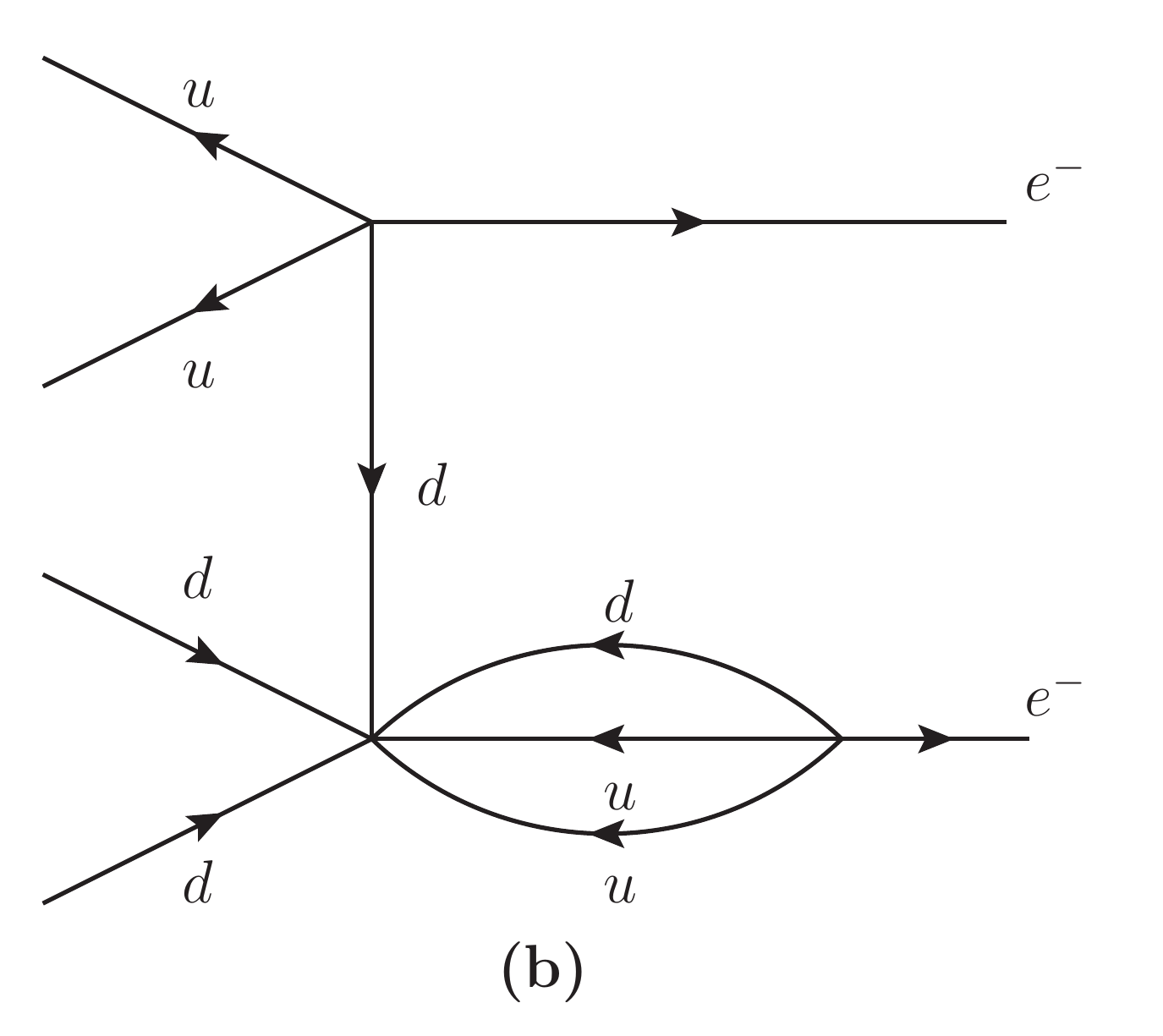}
\vspace*{0.5cm}
\includegraphics[width=4.5cm]{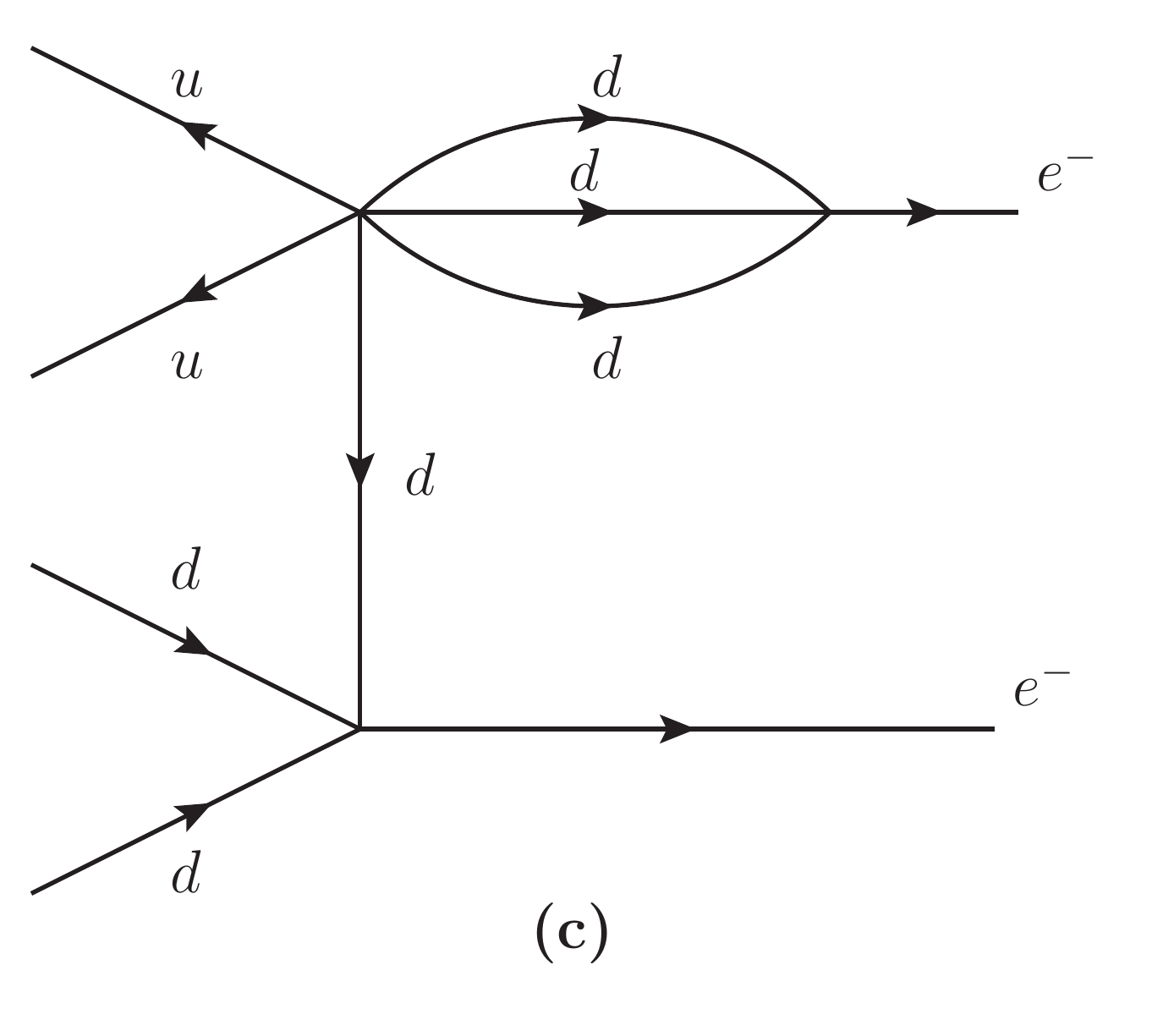}
\caption{Neutrinoless double beta decay diagrams resulting from combining (a) $p \rightarrow e^+ \pi^0$ and $n \rightarrow e^- \pi^+$ decays;
(b) $p \rightarrow e^+ \pi^0$ decay and $n-\bar{n}$ oscillation, and (c) $n \rightarrow e^- \pi^+$ decay and $n-\bar{n}$ oscillation. }
\label{Fig2}
\end{center}
\end{figure}

Let us now suppose that the decays $p \rightarrow e^+ \pi^0$ and $n \rightarrow e^- \pi^+$ are both observed experimentally. By combining the
responsible diagrams of Fig. \ref{Fig1} (a) and Fig. \ref{Fig1} (b), or equivalently the effective Lagrangians of Eqs. (\ref{pdecay}) and (\ref{ndecay}), we can construct a diagram shown in Fig. \ref{Fig2} (a) for neutrinoless double beta decay.  This is the canonical method of determining the Majorana character of the neutrino.  While the effective neutrinoless double beta decay lifetime arising from this diagram would be extremely long, well beyond the experimental reach, observation of $p \rightarrow e^+ \pi^0$ and $n \rightarrow e^- \pi^+$ would nevertheless imply an {\it upper limit} on the double beta decay lifetime, which is sufficient to establish the Majorana character of the neutrino.

We can estimate the lifetime for neutrinoless double beta decay ($\beta\beta_{0\nu})$ arising from Fig. \ref{Fig2} (a) as follows.  In the
amplitude for the standard neutrino mass mediated $\beta\beta_{0\nu}$ diagram, replace $G_F^2$ by  $\Lambda_p^{-2} \left\langle H^0 \right\rangle
\Lambda_n^{-3}$, and $m_\nu^{\rm eff}/\left\langle q^2\right \rangle$ by $1/\left \langle q\right \rangle$ with $\left\langle q\right \rangle \approx 100$ MeV being the average Fermi momentum of the nucleon.  Ignoring any
differences in the nuclear matrix element, and using $m_{\beta \beta} < 0.5$ eV corresponding to the lifetime limit of
$\tau > 2.1 \times 10^{25}$ yrs from GERDA experiment \cite{GERDA}, we obtain an upper limit on the $\beta\beta_{0\nu}$ lifetime arising from Fig. \ref{Fig2} (a) to be
\begin{equation}
\tau_{\beta \beta} < \frac{\tau (p \rightarrow e^+ \pi^0)}{1.4 \times 10^{34}~{\rm yr}} \times \frac{\tau (n \rightarrow e^- \pi^+)}{6.5 \times 10^{31}~
{\rm yr}} \times 10^{113}~{\rm yr}~.
\label{limit}
\end{equation}
While this lifetime is extremely long, it is finite, provided that $\tau (p \rightarrow e^+ \pi^0)$ and $\tau (n \rightarrow e^- \pi^+)$ are
observed and thus finite. No matter how long-lived, it is the finiteness of $\tau_{\beta \beta}$ that is required to establish the Majorana character of the neutrino.

\begin{figure}
\begin{center}
\includegraphics[width=8.5cm]{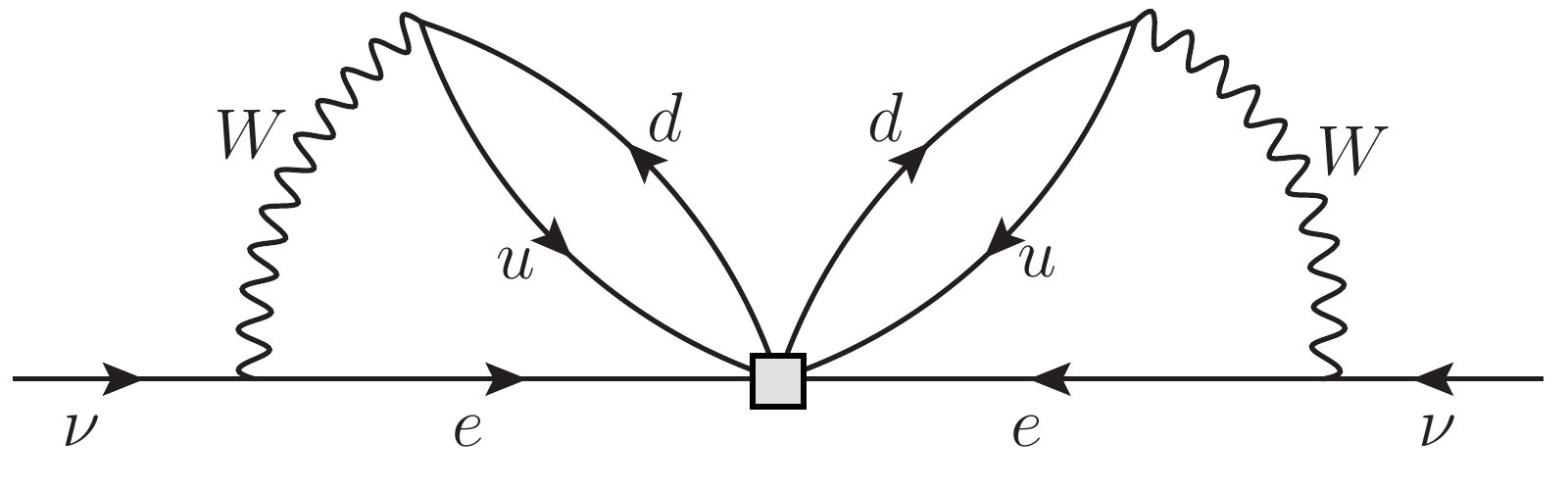}
\caption{ Neutrino Majoran mass induced through any diagram that causes neutrinoless double beta decay.}
\label{Fig3}
\end{center}
\end{figure}

\begin{figure}
\begin{center}
\includegraphics[width=6.0cm]{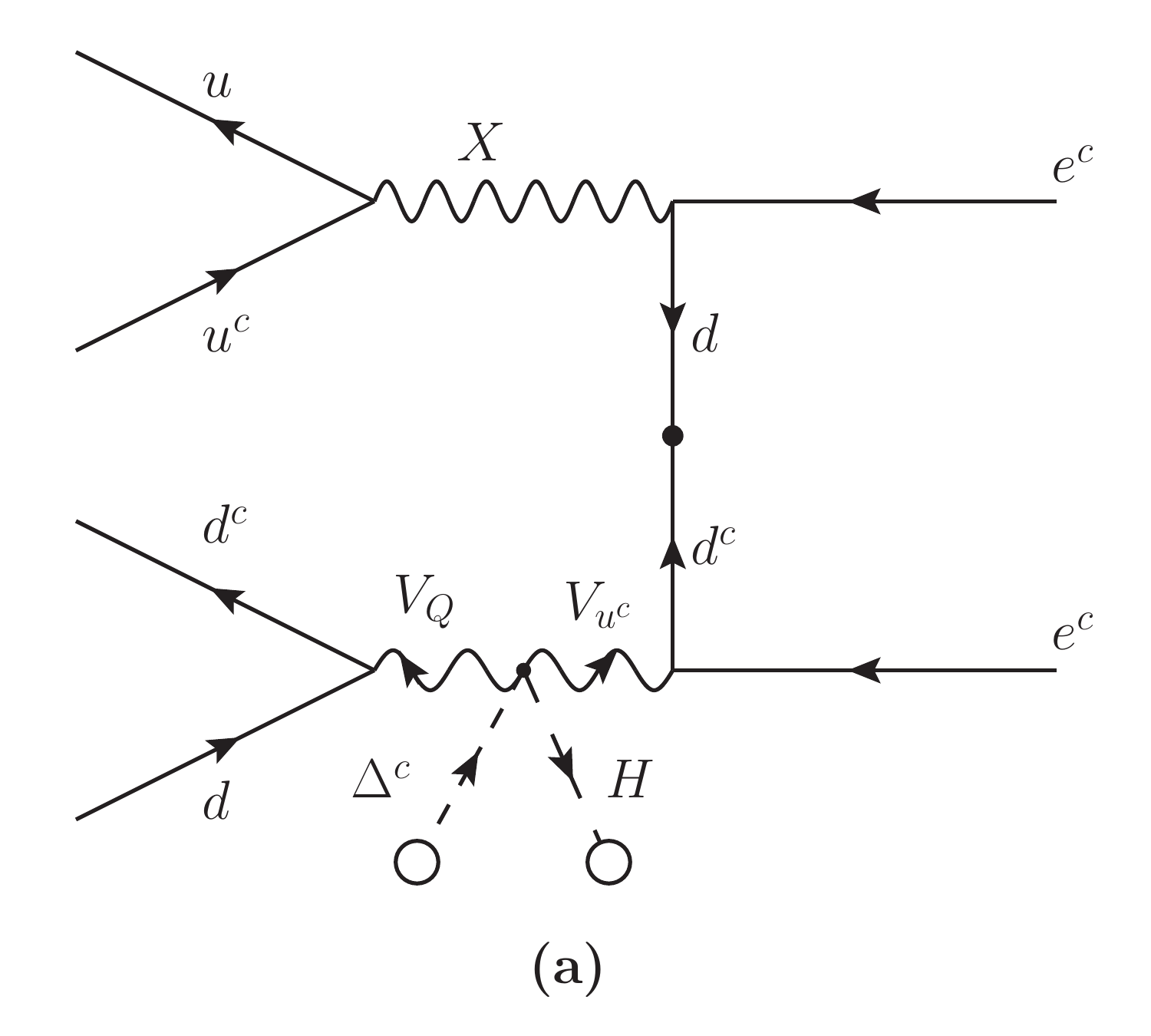}
\hspace*{0.7cm}
\includegraphics[width=6.5cm]{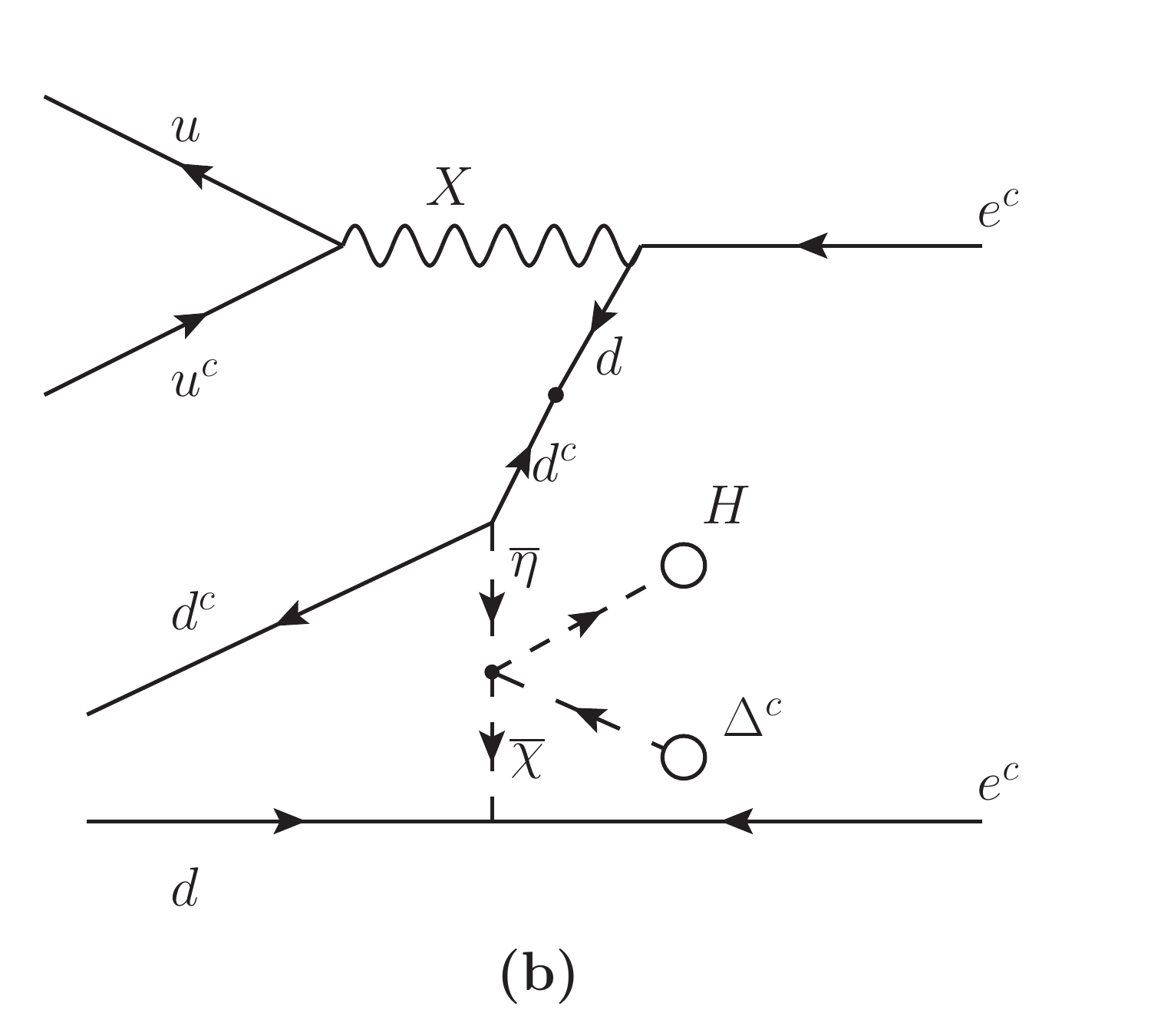}
\caption{ Diagrams inducing neutrinoless double beta decay in $SO(10)$ models by combining $p \rightarrow e^+ \pi^0$ and $n \rightarrow e^- \pi^+$
decays. (a) Exchange of vector bosons of $SO(10)$, and (b) exchange of vector boson and scalars belonging to 120 of $SO(10)$. The quantum numbers
of these fields are $X(3,2,-5/6),\,V_Q(3,2,1/6), \,V_{u^c}(3^*, 1,-2/3), \,\bar{\eta}(3^*,1,-2/3)$ and $\bar{\chi}(3^*, 2, -7/6)$
under Standard Model gauge symmetry. Here $H(1,2,1/2)$ is the SM Higgs field, and $\Delta^c$ is the Higgs field responsible for $B-L$ breaking \cite{bm1}. }
\label{Fig4}
\end{center}
\end{figure}

The diagram of Fig. \ref{Fig2} (a) would generate a small Majorana neutrino mass through the loop diagram of Fig. \ref{Fig3}.
This diagram exists in all models where $\beta\beta_{0\nu}$ occurs \cite{valle}.  The box at the center of the diagram stands for
the fundamental vertex leading to $\beta\beta_{0\nu}$. The induced neutrino mass, while extremely tiny,
is nonzero, which is sufficient to establish the Majorana character of the neutrino.  Evaluation of Fig. \ref{Fig3} naively would
be fraught with divergencies, but this is not the case once the box is opened up.  (For a general discussion of evaluation of this diagram
see Ref. \cite{lindner}.)  We focus on a well motivated and realistic scenario where
both $p \rightarrow e^+ \pi^0$ and $n\rightarrow e^- \pi^+$ decays are potentially near the current experimental limit, viz.,
non-supersymmetric $SO(10)$ grand unified theories with an intermediate scale \cite{bm1}.  The diagram for $\beta \beta_{0\nu}$ of
Fig. \ref{Fig2} (a) can arise in these models through the diagrams shown in Fig. \ref{Fig4}.  Fig. \ref{Fig4} (a) shows these processes
mediated by $SO(10)$ gauge bosons, and Fig. \ref{Fig4} (b) by scalar boson belonging to the 120 of Higgs.  In Fig. \ref{Fig4} (b) if the scalars
fields have masses of order $10^{11}$ GeV, which may be identified as the intermediate symmetry breaking scale of $SO(10)$, then
$n \rightarrow e^- \pi^+$ decay may be in the observable range.  $p \rightarrow e^+ \pi^0$ decay is mediated by the $X$ gauge bosons,
and may also be within reach.  If we insert one of the figures of Fig. \ref{Fig4} in place of the box of Fig. \ref{Fig3}, we obtain finite
and small neutrino Majorana masses. We estimate the induced mass to be
\begin{equation}
m_\nu \sim \frac{g^4}{(16 \pi^2)^4} \,
 \frac{m_e^2 \, m_d^2\, m_u \left\langle H^0 \right\rangle}{\Lambda_p^2\,  \Lambda_n^3}~.
\end{equation}
Numerically, $m_\nu \sim 5 \times 10^{-86}$ GeV, if $\Lambda_p$ and $\Lambda_n$ are at their current experimental limits.  We stress again
that it is not the numerical value of $m_\nu$ that is important here, but the fact that $m_\nu$ is nonzero.  Obviously, neutrino oscillation
phenomenology would require other sources of Majorana mass generation, which is present in $SO(10)$ models, in the form of the
seesaw mechanism.  

Now let us turn to the case where $p \rightarrow e^+ \pi^0$ and $n-\bar{n}$ oscillations are observed.
In this case, the diagram shown in
Fig. \ref{Fig2} (b) would lead to neutrinoless double beta decay.  Note that in the bottom half of this diagram, a $n-\bar{n}$ vertex has been
combined with a $p \rightarrow e^+ \pi^0$ vertex to generate effectively a $n \rightarrow e^- \pi^+$ vertex. Thus this case would lead to
identical results for the Majorana nature of the neutrino as in the case of $p \rightarrow e^+ \pi^0$ and $n \rightarrow e^- \pi^+$ discussed earlier.
Due to the appearance of two loops, the lifetime for $\beta\beta_{0\nu}$ arising from Fig. \ref{Fig2} (b) would be about four orders of
magnitude longer than the one quoted in Eq. (\ref{limit}).  The induced neutrino Majorana mass would be smaller, of order
$10^{-90}$ GeV.

\begin{figure}
\begin{center}
\includegraphics[width=8.5cm]{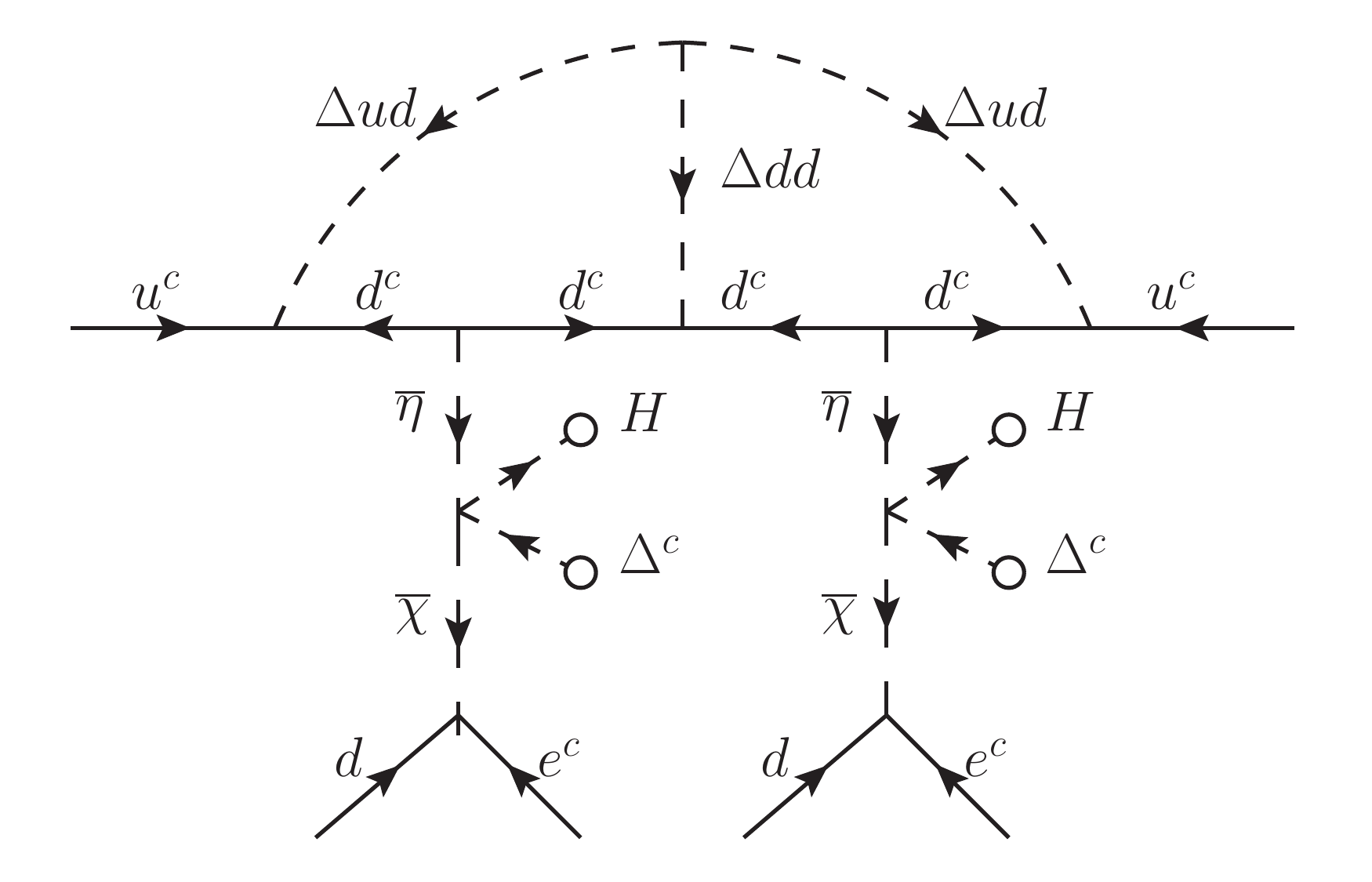}
\caption{Leading diagram for neutrinoless double beta decay in a model with observable $n \rightarrow e^- \pi^+$ decay and $n-\bar{n}$ oscillations.
   Here $\Delta_{ud}$ and $\Delta_{dd}$ are color sextet scalars.} \label{Fig5}
\end{center}
\end{figure}

The third scenario assumes that $n \rightarrow e^- \pi^+$ decay and $n-\bar{n}$ oscillations are observed.  Note that both of these
processes change $B-L$ by two units.  One can construct a $\Delta(B-L) = 0$ effective operator by combining $n \rightarrow e^- \pi^+$ decay
diagram with the conjugate of $n-\bar{n}$ oscillation diagram, as shown in Fig. \ref{Fig2} (c).  This diagram shows how neutrinoless double
beta decay occurs in this case. A simple way of generating observable $n-\bar{n}$ oscillations is through the exchange of color sextet scalars
which are present in gauge theories based on $SU(2)_L \times SU(2)_R \times SU(4)_c$ \cite{marshak} as well as in $SO(10)$.  
As noted already, the existence of
color triplet scalars $\bar{\eta}$ and $\bar{\chi}$ from the 120 Higgs of $SO(10)$ at an intermediate scale of $10^{11}$ GeV or lower could lead
to observable $n \rightarrow e^- \pi^+$ decay.  When these two sources are combined, we find the dominant diagram for neutrinoless double
beta decay to be the one shown in Fig. \ref{Fig5}. This diagram is structurally identical to the two-loop neutrino mass generation diagram of the
model of Ref. \cite{zee}, which has been studied in detail.  Using the results of Ref. \cite{cosmin}, we obtain the neutrinoless double beta
decay amplitude to be
\begin{equation}
A_{\beta \beta}^{0\nu} \simeq \frac{\hat{I}}{(16 \pi^2)^2} \frac{M_{dd}^2 M_{ud}^2}{\Lambda_{n\bar{n}}^5} \times \frac{\left\langle H^0 \right\rangle^2}
{\Lambda_n^6}
\end{equation}
where $\hat{I}$ is a dimensionless integral of order unity \cite{cosmin}.
Using the present lower limits on $\Lambda_n$ and $\Lambda_{n\bar{n}}$, we obtain the lifetime for neutrinoless double beta decay
to be of order $10^{131}$ yrs in this case.  Thus we see that in all three cases the Majorana character of the neutrino will be established.

\begin{figure}
\begin{center}
\includegraphics[width=6.5cm]{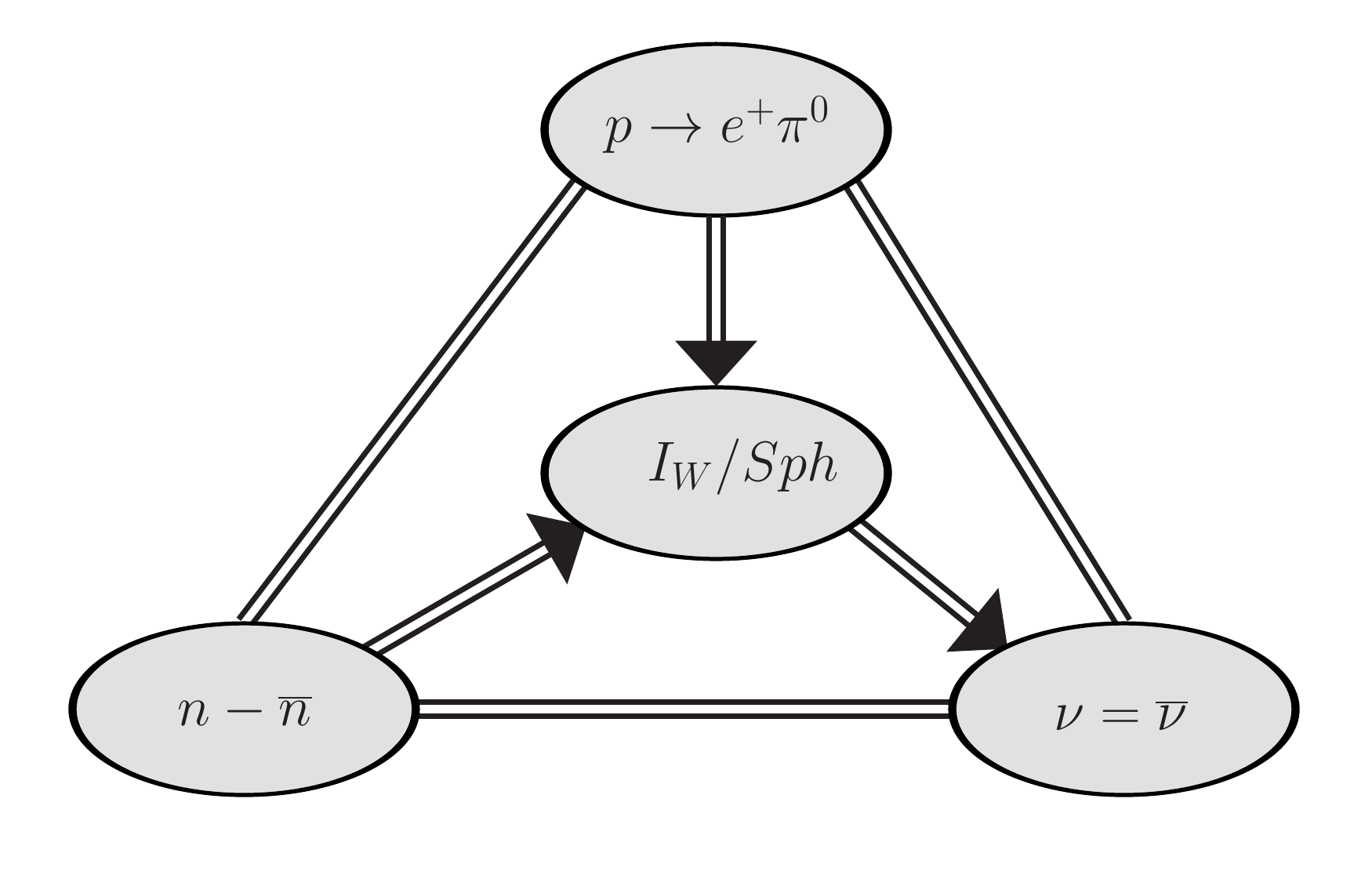}
\caption{ ``$B-L$ Triangle'' explains how discovering $n-\bar{n}$ oscillations and proton decay would impliy that neutrinos are Majorana fermions} \label{Fig6}
\end{center}
\end{figure}

We conclude by making an observation that connects baryon number violation and neutrino Majorana mass through the non-perturbative
instanton/sphaleron configurations of weak interactions. These solutions lead to an effective operator involving the twelve Standard Model
doublet fermions $QQQQQQQQQLLL$, with an exponentially suppressed coefficient \cite{thooft} (at zero temperature) which is very difficult to
observe at colliders.  Here $Q$ and $L$ are the quark and lepton fields. When expanded, after rotations to bring the mass matrices to diagonal
forms, this term would contain terms which look like $uddudd uude \nu\nu$.  We can rewrite this as a product of three parts: $[uddudd] \cdot [uude]\cdot [\nu\nu]$. Note that the first part is the piece that contributes to $n-\bar{n}$ oscillations, the second part is relevant to $p\to e^+\pi^0$ decay, and the last part leads to Majorana mass for the neutrinos. One can represent this in terms a triangle shown in Fig. \ref{Fig6}. Thus, if $p\to e^+\pi^0$ and $n-\bar{n}$ oscillations are observed, the instanton/sphaleron configuration would imply that neutrino has a Majorana mass. This time it is a direct induction of Majorana mass rather than neutrinoless double beta decay unlike the cases discussed earlier. We realize that the weak instanton/sphaleron effects have not been directly observed and in that sense, this connection is not based on purely experimental observations. Nonetheless we believe that this is an
interesting theoretical connection.

The work of KSB is supported in part by the US Department of Energy Grant No. de-sc0010108 and  RNM is supported
in part by the National Science Foundation Grant No. PHY-1315155.  We thank Saki Khan for graphical help.

\end{document}